\documentstyle[psfig,epsf,conf-X]{article}
\begin{document} 
\small
\heading{%
%
Thermal equilibrium of cold clouds in PKS 0745-191
\footnote{Poster presented for the International Workshop {\it Large scale 
structure in the X-Ray Universe}, Santorini Island (Greece), September 20-22$^{th}$, 1999}
}
\par\medskip\noindent
\author{%
Lukas Grenacher$^{1,2}$, Philippe Jetzer$^{1,2}$ and Denis Puy$^{1,2}$
}
\address{%
Paul Scherrer Institute, Laboratory for Astrophysics, 5232 Villigen and
}
\address{%
Institute of Theoretical Physics, University of Zurich, 8057 Zurich, Switzerland
}
%

\begin{abstract}
PKS 0745-191 is a powerful radio source with one of the largest known 
cooling flows \cite{degrandi}. A fraction of the 
cool gas could accumulate into small molecular clouds. 
We studied the minimum 
temperature which can be reached by sub-clouds (resulting from the 
fragmentation of bigger clouds) in the cooling flow region.
\end{abstract}
\section{Introduction}
Molecules such as $H_2$ and $HD$ are expected to be present in the 
post-recombination gas and due to their cooling properties they 
can thermally influence the gravitationnal collapse of the first objects
which formed in the Universe \cite{puy1}. At low temperatures, these 
molecules with some traces of $CO$ could also be 
present in the intracluster gas, where they could act as important coolant 
in cooling flows.\\
In this chemically simple gas the molecules are mainly excited collisionally.
Followed by a radiative de-excitation in the optically thin medium this
leads to an energy loss for the gas clouds and thus to a cooling.\\ 
The aim of this communication is to discuss the minimum temperature 
achievable by clouds located in the region of the cooling flow of 
PKS 0745-191.

\section{Equilibrium distance}
We have computed the molecular cooling (including radiative transfer effects)
due to $H_2$, $HD$ and $CO$ for small clouds which are the result of a 
fragmentation process of bigger clouds in cooling flows \cite{puy2}.\\
In our calculation we included also an attenuation factor $\tau$ which
caracterizes the column density surrounding the sub-clouds.
Thus the attenuated bremsstrahlung flux coming from the 
intracluster gas heats the clouds located in the cooling flow at 
a distance $r$ from the cluster center, and so thermal balance between 
heating and cooling defines an equilibrium temperature of the sub-clouds 
at a distance $r=R_{eq}$ inside the cooling flow region 
(i.e. $R_{eq}<r_{cool}$).\\
\newline
The following column densities are adopted for a typical small cloud (with
$n_{H_2}=10^6$cm$^{-3}$ and the orto-para ratio equal to 1):\\
$N_{CO}=10^{14}\, {\rm cm}^{-2} \, \, {\rm and} \, \,  
N_{H_2}=2 \times 10^{18} \, {\rm cm}^{-2}$,
which corresponds to a $CO$ abundance:
$\eta_{CO}\sim 5 \times 10^{-5}$. For $HD$  instead we assume the 
primordial ratio $\eta_{HD} \sim 7 \times 10^{-5}$.\\
\newline
In Figure 1 we plotted the equilibrium temperature of clumps at the 
equilibrium distance $R_{eq}$ for different values of the attenuation factor 
$\tau$. 
We see that 
low equilibrium temperatures are achieved at distances smaller than
the cooling 
radius $r_{cool}$.
\begin{figure}[h]
\centerline{\vbox{
\psfig{figure=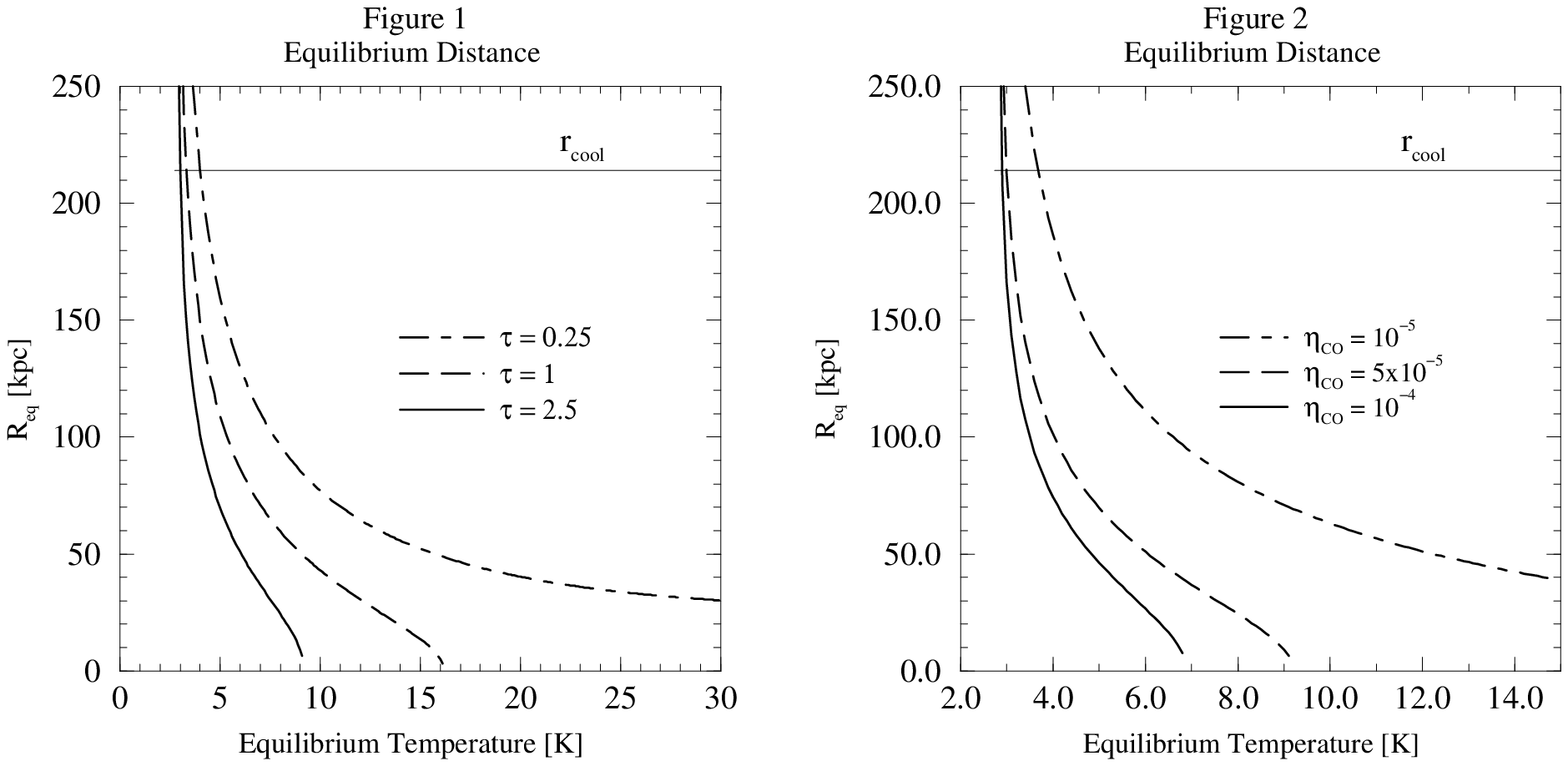,height=9cm,angle=-90}
}}
\end{figure}
In Figure 2 we plotted the equilibrium distance as a function of the 
equilibrium 
temperature for different values of $\eta_{CO}$ and $\tau=2.5$ is 
kept fixed.
Indeed, the $CO$ abundance and thus $\eta_{CO}$ is an important 
parameter which is, however, not well known.\\
\newline
We thus find that a fraction of the gas in the cooling flow of
PKS 0745-191 could be very cold, which might form small clouds
via fragmentation.

\acknowledgements{We would like to thank M. Plionis and I. 
Georgantopoulos for organizing this pleasant conference. This work has been 
supported by the {\it Dr Tomalla Foundation} and by the Swiss NSF.}

\vfill

\begin{iapbib}{99}{

\bibitem{degrandi} De Grandi S., Molendi S., 1999 A\&A Lett. in press 


\bibitem{puy1} Puy D., Alecian G., Le Bourlot J., Leorat J., Pineau des 
For\^ets G., 1993 \aeta 267, 337



\bibitem{puy2} Puy D., Grenacher L., Ph. Jetzer, 1999, \aeta 345, 723
}
\end{iapbib}
\end{document}